\documentclass{aa}
\usepackage{amsmath,graphicx,amssymb}
\usepackage{txfonts}
\usepackage{natbib}
\bibpunct{(}{)}{;}{a}{}{,}

\newcommand\hvezda{\object{HR~7224}}
\newcommand{\zav}[1]{\left(#1\right)}
\newcommand{\hzav}[1]{\left[#1\right]}
\newcommand\intvidpo{\!\!\int\limits_{\begin{array}{c}\text{\scriptsize
visible}\\[-2mm]\text{\scriptsize surface}\end{array}}\!\!}

\newlength\staretab

\makeatletter
\def\sgn{\mathop{\operator@font sgn}\nolimits}
\makeatother

\begin{document}

\title{The nature of the light variability of the silicon star HR 7224}

\author{J.~Krti\v{c}ka\inst{1} \and Z.~Mikul\'a\v sek\inst{1,2}
        \and G. W. Henry\inst{3} \and J.~Zverko\inst{4}
        \and J.~\v Zi\v z\v novsk\'y\inst{4}\and
        J. Skalick\'y\inst{1}\and
        P.~Zv\v e\v rina\inst{1}}

\offprints{J.~Krti\v{c}ka,\\  \email{krticka@physics.muni.cz}}

\institute{Department of Theoretical Physics and Astrophysics,
           Masaryk University, Kotl\'a\v rsk\' a 2, CZ-611\,37 Brno, Czech Republic
            \and
            Observatory and Planetarium of J. Palisa, V\v SB -- Technical
            University, Ostrava, Czech Republic
            \and
            Center of Excellence in Information Systems, Tennessee
            State University, Nashville, Tennessee, USA
            \and
            Astronomical Institute, Slovak Academy of Sciences,
            Tatransk\'{a} Lomnica, Slovak Republic
            }

\date{Received}

\abstract {Although photometric variations of chemically peculiar (CP) stars are
frequently used 
to determine
their rotational periods, the detailed
mechanism of their light variability remains poorly understood.} {We simulate
the light variability of the star HR~7224 using the observed surface
distribution of silicon and iron.} {We 
used
the TLUSTY model atmospheres
calculated for the appropriate silicon and iron abundances to obtain the
emergent flux and to predict the rotationally modulated light curve of the star.
We also 
obtained
additional photometric measurements and 
employed
our own regression
procedure to derive a more precise estimate of the light elements.} {We show
that the light variation of the star can be explained as a result of i) the
uneven surface distribution of the elements, ii) the flux redistribution from
the ultraviolet to the visible part of the spectrum, and iii) rotation of the
star. We show that the silicon bound-free transitions and iron bound-bound
transitions provide the main contribution to the flux redistribution, 
although
an
additional source of opacity is needed. We confirm that numerous iron lines
significantly contribute to the well-known depression at $5200\,$\AA\ and
discuss the connection between iron abundance and the value of peculiarity index
$a$.} {The uneven surface distribution of silicon and iron is able to explain
most of the rotationally modulated light variation in the star HR~7224.}


\titlerunning{The nature of the light variability of the Si star
HR~7224}
\authorrunning{J.~Krti\v{c}ka et al.}
\maketitle

\section{Introduction}

Chemically peculiar (CP) stars represent 
a large
class of upper main
sequence stars where the processes of radiative diffusion and gravitational
settling in their atmospheres give rise to pronounced deviations in the chemical
composition of these stars from the solar value (\citealt{vaupreh},
\citealt{mpoprad}). 
The CP stars are natural laboratories for testing modern model atmospheres
thanks to the unusual chemistry with rather strong under- or overabundance of
some elements.
The application of advanced modelling
techniques, such as model atmospheres with magnetic fields (e.g.,
\citealt{malablaj}, \citealt{malablat}), radiative diffusion codes, and Doppler
imaging techniques \citep{choch}, provides us with detailed information about the
surface structure of these stars. Despite these fascinating advances in their
study during recent decades, the light variability of CP stars is still poorly
understood.

Some CP stars show inhomogeneous surface distribution of chemical elements on
their surface \citep[e.g.,][]{choch} as determined from rotationally modulated
spectral line variability \citep[see, e.g.,][]{leh1}. The uneven surface
distribution of various elements, together with rotation, has been presumed to
be the 
origin
of these stars' light variability. However, the details of
this mechanism have not been determined. Line blanketing by multiple lines of
overabundant elements (mainly iron) and the flux redistribution induced by these
lines has been proposed as one of the causes of the light variability
\citep{molnar}. Other mechanisms proposed include the influence of bound-free
transitions \citep{peter,lanko}, surface temperature differences or variable
temperature gradients \citep{koda,biltep}, and the presence of magnetic fields
\citep[e.g.,][]{malablaj}. Finally, circumstellar matter, if present, may also
influence the light curves \citep{labor,nakaji,smigro,towog}.

One of the first attemps to simulate the light variability of CP stars was done
by \citet{krivo},
who reproduced the light curve of the CP
star \object{CU~Vir}. However, detailed modelling of CP star light variability
had to await precise model atmospheres \citep[e.g.,][]{bstar2006}, improved
opacity data \citep{topt}, and much faster computers. \citet{myhd37776} took
advantage of these tools and used the surface maps of \citet{choch} to simulate
the light curve of \object{HD 37776} successfully.
They 
demonstrate
that the
inhomogeneous surface distribution of silicon and helium, along with the
bound-free transitions of these elements, accounted for most of the light
variability in this star.

Much work remains to be done to understand 
the
light variability in CP stars. In
particular, the role of iron, which is found to be significantly underabundant
in the atmosphere of \object{HD 37776} \citep{choch}, remains to be clarified.
For this purpose we selected 
the silicon star \hvezda, whose silicon and iron surface
distributions were derived via Doppler imaging by \citet{leh2}.

\section{CP star \hvezda}


\hvezda~(HD 177410, HIP 93187, EE~Dra) is a rapidly rotating CP star
classified as A0p Si by \citet{cow}. \citet{mlyn} classified
\hvezda\ as a B9.5 IIIp Si star, based on low-resolution spectroscopy.
\citet{leh1,leh2} found that the star has enhanced Si and Fe
abundances, whereas He is extremely depleted; they classified it as
a helium-weak silicon CP star.

\citet{winzer} found photometric variability in \hvezda\ and determined
a period of 1\fd1663 from 48 $\mathit{UBV}$ observations done in
1970--72. \citet{adela} derived a new period of 1\fd123095 based on
616 Str\"omgren $\mathit{uvby}$ observations taken in the 1993--94 and
1994--95 observing seasons with the Four College Automated Photometric
Telescope (FCAPT) along with the older $V$ measurements of \citet{winzer}.
\emph{Hipparcos} obtained 409 $\mathit{Hp}$, $B_{\mathrm{T}}$, and
$V_{\rm{T}}$ observations in 1989--92 \citep{esa97}, and \citet{esa98}
determined the photometric period to be 1\fd123248.

\begin{figure}[t]
\centering \resizebox{1.0\hsize}{!}{\includegraphics{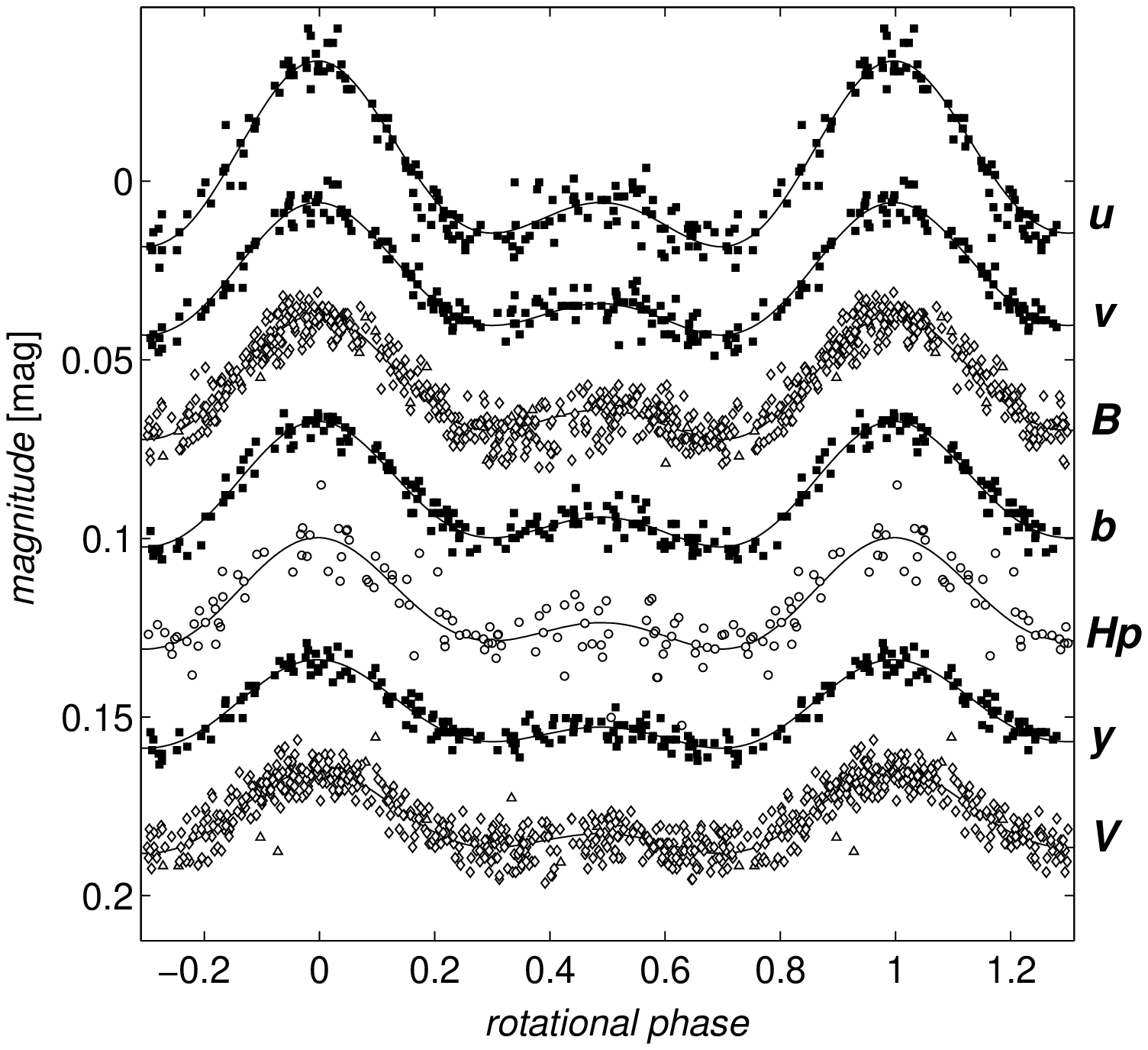}}
\caption{Photometric observations of \hvezda\ plotted as a function
of the new linear phase in Eq.~\eqref{linef}. Data are from \citet
{winzer}, $\mathit{BV}~(\triangle)$; \citet{adela}, $\mathit{uvby}~(\Box)$;
Hipparcos \citep{esa97}, $\mathit{Hp}$ (o), and the T3 APT (this paper),
$\mathit{BV}~(\diamondsuit)$.  Solid lines denote the fit according to
Eq.~\eqref{regmod}.
$B_{\text{T}}V_{\text{T}}$ and $U$ magnitudes are not plotted here due to
their large scatter.}.
\label{curves}
\end{figure}

All of the optical light curves phase together fairly well and show a
double-wave with two maxima of different height (see Fig. \ref{curves}).
\hvezda~was recently classified from the shape of its light curve as a prototype
of a photometrically simple CP star with a double-wave light curve with two
unequally prominent bright spots centred on the phases $\varphi=0.0$ and 0.5
\citep{simply}. There are only two attempts to measure the magnetic field of
\hvezda\ available in the literature \citep{bolek,leh2}; both produced negative
results. 
Recent
magnetic field measurements (Kudryavtsev, private
communication) also gave negative results, however the observations are still
continuing
to cover the whole rotational cycle.

\citet{adelc} reported an unprecedented change in the photometric behaviour of
\hvezda. Comparing time series observations taken with FCAPT before 1996 with
those taken in 2003, he found that the amplitude of variability had increased
from a typical value of 0.04 mag to 0.21 mag.  He also reported that the period
of variability had changed from 1\fd123 to 101 days. Unfortunately, the
photometric data from this critical era have not been published.

Adelman's astonishing result, along with the lack of available
high-resolution spectroscopy of \hvezda, motivated \citet{leh1} to
carry out an extensive spectroscopic observing program on the star.
Their 564 high-resolution spectrograms allow determination of radial
velocities via cross correlation to an accuracy better than 100
m\,s$^{-1}$. They found radial velocity variations in this swiftly
rotating star with an amplitude of 15\,km\,s$^{-1}$ and a period of
1\fd123248(9), in excellent agreement with the previous
determination by \citet{esa98}. They found no further periodicities,
in particular nothing around the 101-day period of \citet{adelc}.

\citet{leh2} used the same 564 spectrograms to derive Doppler images of
surface elemental distributions.  Their map of the silicon and iron abundance
on the surface of \hvezda\ is an ideal starting point for the simulation
of the star's expected light curves. To avoid any difficulties phasing
the spectroscopic and photometric observations together, we acquired new
photometric observations to improve the rotational ephemeris of \hvezda.

\subsection{New BV photometry of \hvezda}

Our new $BV$ photometry was acquired between March and May 2008 with
the T3 0.4\,m automatic photoelectric telescope (APT) at Fairborn
Observatory. This APT uses a temperature-stabilised EMI 9924B
photomultiplier tube to measure photon count rates through Johnson
$B$ and $V$ filters. \hvezda\ and its comparison stars were measured
in the following sequence, termed a group observation:
\texttt{K-S-C-V-C-V-C-V-C-S-K}, where \texttt{K} is the check star
(HD\,172728, $V=5.74$, $B-V=-0.045$, A0\,V), \texttt{C} is the
comparison star (HD\,172569, $V=6.07$, $B-V=0.279$, F0\,V),
\texttt{V}~is \hvezda, and \texttt{S} is a sky measurement. Three
\texttt{V-C} and two \texttt{K-C} differential magnitudes are formed
from each sequence and averaged together to create group mean
differential magnitudes.  The typical precision of a group mean is
0.003--0.004\,mag for this telescope.  To filter out observations
taken in non-photometric conditions, group means with a standard
deviation greater than 0.01 mag were discarded.  From the 572 group
observations collected by the telescope, 478 $B$ and 462 $V$ group
means survived the filtering process and were used in this analysis.

The surviving group means were corrected for differential extinction
with nightly extinction coefficients, transformed to the Johnson system
with yearly-mean transformation coefficients, and treated as single
observations thereafter. Up to five group observations were acquired
each clear night at intervals of approximately two hours. Further
information on the operation of the APT and the analysis of the data
can be found in \citet{h95a,h95b} and \citet{ehf03}.

The photometric data used in this analysis are available through
SIMBAD and the {\it On-line database of photometric observations of
mCP stars} (http://astro.physics.muni.cz/mcpod).

\subsection{New linear ephemeris}

To derive the ephemeris, we used all available photometric data on
\hvezda\, including the $\mathit{UBV}$ measurements of \citet{winzer},
the $H\!p$ and $B_{\text{T}}V_{\text{T}}$ observations from Hipparcos
\citep{esa97}, the Str\"omgren $\mathit{uvby}$ photometry by
\citet{adela}, and the T3 APT $\mathit{BV}$ observations in this paper.
The total of 2013 individual photometric measurements covers a time
span of 38 years; they were supplemented with the 564 measurements of
silicon line equivalent widths ($\it{EW}$) obtained by \citet{leh2} in
2004--05 and kindly provided to us by Dr.~H.~Lehmann.

The analysis of these data to determine the ephemeris is described in
\citet{brzda}. It is assumed that all the photometric and spectroscopic
variability can be expressed in terms a simple regression model.

\subsubsection{Regression model and robust regression procedure}

The formulation of the observed phase variation of the light and equivalent
widths of silicon lines in terms of a general regression model is relatively
simple since the light and $\it{EW}$ phase curves are similar in shape, though
shifts may exist in some cases.  Consequently, we can express them with the
following relations: \begin{equation} y_{cj}(t_i)= \overline{y}_{cj}
+\textstyle\frac{1}{2}\,A_c\,F(\vartheta_{ij}),
\quad\vartheta_{ij}=\zav{t_i-M_0-\Delta M_{0j}}/P, \label{regmod} \end{equation}
where $y_{cj}$ is the predicted value of an observed quantity (magnitude,
$\it{EW}$) in a colour (pass band) $c$ (in this case
$\mathit{u,\,U,\,v,\,B,\,b,\,Hp,\,y,\,V}$, and $\mathit{EW}$) of the $j$-th set
of measurements (Winzer, Hipparcos, Adelman I and II, Lehmann, and this paper)
at the particular JD$_{\mathrm{hel}}$ instant $t_i$, and $\overline{y}_{cj}$ is
the weighted-mean value of the quantity. Note that observations in $B_\text{T}$
were included into $B$ and $V_\text{T}$ into $V$. Here $A_c$ is an ``effective
amplitude", the parameter robustly describing the measure of the variability in
the colour $c$, and $F(\vartheta)$ is a normalised (unique) function expressing
the form of the observed phase curves \citep[for details see][]{mikdata,mikzoo}.
It is a periodic function of the monotonically growing ``phase function"
$\vartheta$ introduced by \citet{brzda}

The phase function of time $\vartheta(t)$ is determined by the
ephemeris, which is given in the second part of Eq.\,\eqref{regmod}.
We assume that the period $P$ of \hvezda\ is constant and define zero
phase to be the light maximum $M_0$ that is situated nearest to the
weighted centre of gravity of measurements used for the ephemeris
determination. $\Delta M_{0j}$ are free parameters correcting $M_0$ for
some groups of observation. We applied them only when justified,
particularly for observations of \citet{winzer} (see the following
discussion) and for the equivalent widths of the Si lines.

Function $F(\vartheta)$ is the simplest normalised
periodic function that represents the observed photometric and
spectroscopic variations of \hvezda\ in detail.
The phase of maximum brightness
is defined to be 0.0, and the amplitude is defined to be 1.0. The
function, being the sum of three orthogonal terms, is described by
two parameters $a_1$ and $a_2$. The first parameter quantifies the
symmetric portion of the deviation of the light curve from a simple
cosine course; the second parameter expresses any asymmetry in the
light curve:
\begin{eqnarray}\label{fce}
 F(\vartheta,\,
a_1,\,a_2)=\sqrt{1\!-\!a_1^2\!-\!a_2^2}\ \cos(2\,\pi\,\vartheta)+
a_1\cos(4\,\pi\,\vartheta)\nonumber \\
+\,a_2\hzav{\textstyle{\frac{2}{\sqrt{5}}}\,\sin(2\,\pi\,\vartheta)
-\textstyle{\frac{1}{\sqrt{5}}}\sin(4\,\pi\, \vartheta)}.
\end{eqnarray}

\begin{figure}[t]
\centering \resizebox{0.9\hsize}{!}{\includegraphics{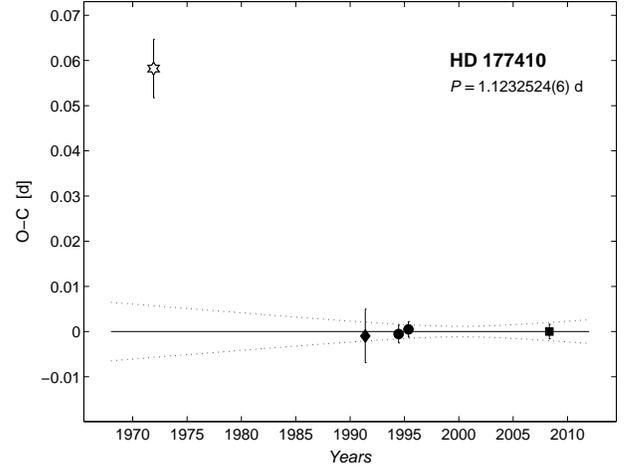}}
\caption{O-C diagram of \hvezda\ computed with the new linear
ephemeris in Eq.~\eqref{linef}. The open star represents the observations of
Winzer, the filled diamond corresponds to the Hipparcos measurements,
the filled circles are from the two seasons of observations by Adelman,
and the filled square gives the result of the BV observations in this
paper.  The solid line denotes the line of constant period and the dotted
line is the uncertainty.}
\label{OC}
\end{figure}

\subsubsection{New findings}
We used the same robust, iterative, regression procedure described in
Sect.~3.3 of \citet{brzda} to determine a new linear ephemeris for
\hvezda.  The ephemeris consists of the period $P$, the JD$_{\rm hel}$
of the light curve maximum $M_0$, the mean shift of Winzer's
measurements $\Delta M_{0\mathrm{win}}$, and the mean shift of the
maximum of the Si equivalent width, $\Delta M_{0\mathrm{EW}}$, relative
to the light-curve maxima:
\begin{align}\label{linef}
 &P=1\fd1232524(6),&M_0=2\,451\,582.7164(12), \\
 &\Delta M_{0\mathrm{win}}=0\fd058(10),
 &\Delta M_{0\mathrm{EW}}=0\fd0341(32). \nonumber
\end{align}

\begin{table}[ht]
\caption{O-C derived from photometry and spectroscopy of \hvezda.
$s$ denotes standard deviation in mmags.} \label{OCtab} \centering
\begin{tabular}{rrlrcl}
\hline \hline \rule{0pt}{1.07em}
Year~&
O-C [d] \hspace{3pt}
&$~s$&$N~$&Data&Source\\
 \hline
  1972 &   0.0580(70)  &  4.4  &  48 & $\it{UVB}$                     & \citet{winzer}\\
  1991 & $-0.0010(60)$ & 11    & 409 & $B_{\text{T}}V_{\text{T}}H\!p$ & \citet{esa97}\\
  1994 & $-0.0005(20)$ & $3.8$ & 317 &$uvby$                          & \citet{adela}\\
  1995 &   0.0005(18)  & $3.3$ & 299 &$uvby$                          & \citet{adela}\\
  2006 &   0.0341(24)  &       & 564 &$EW$                            & Lehmann et al.\\
  2008 &   0.0000(17)  & $4.8$ & 940 &$\mathit{BV}$                   & this paper\\
\hline
\end{tabular}
\label{table}
\end{table}

Our period agrees with those determined by \citet{esa98} and
\citet{leh1} within their uncertainties; the period appears to be
constant over the past 17 years (see Fig.~\ref{OC} and Table~\ref{OCtab}).
The nonzero shift $\Delta M_{0\mathrm{EW}}$ is likely connected with the
fact that uneven horizontal distribution of silicon is not a unique 
source of the light variability (see Sect.~\ref{kapsrov}).
Winzer's (1974) $\mathit{UBV}$ light curves have the same
shape as the more recent ones but are shifted with respect to the
new ephemeris by $0\fd058(10)$ (6\,$\sigma$). Accepting the
reality of the shift, we can speculate about the possibility of a
transient rotational braking that might have occurred before the
Hipparcos mission. We discussed a similar event in the He-strong
star HD~37776 \citep{brzda}. In the case of \hvezda\, however,
we have no observational evidence of such deceleration between the
Winzer and Hipparchos epochs. If we assume a constant rate
of deceleration $(\dot{P}=\text{const.})$, then we need
$\dot{P}\simeq3.4\times10^{-9}$ in order to explain the of O-C
shift of Winzer's observations. Such a deceleration would manifest
itself in the lengthening of the period $P$ by 1.8 second during
this time interval of 17 years (1972 -- 1989).

\begin{figure}[t]
\centering
\resizebox{0.80\hsize}{!}{\includegraphics{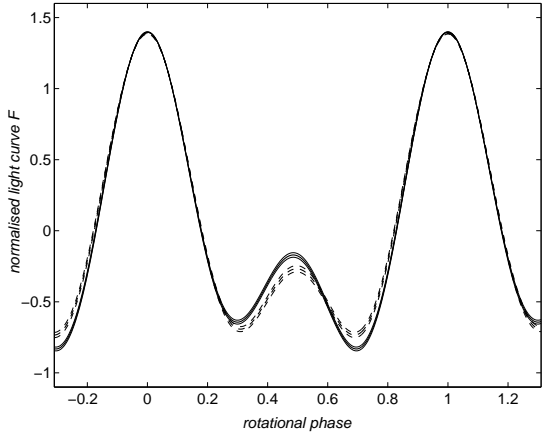}}
\caption{Shapes of the normalised light curves of \hvezda\ before 2000
(solid lines) and after 2000 (dashed lines). The outer light lines
demarcate the uncertainty in the course of the light curves.}
\label{ChLC}
\end{figure}

The parameters $a_1,\,a_2$, which describe the shape of the
normalised function $F(\vartheta,\,a_1,\,a_2)$
(Eq.\,\eqref{fce}) show some minor differences before and after the year 2000:
\begin{align}
\text{before}\ 2000\quad\quad &a_1=0.610(10);\quad &a_2=0.084(12);
\nonumber\\
\quad\text{after}\ 2000\quad\quad &a_1=0.560(13);\quad
&a_2=0.018(15).
\end{align}
If real, this means that the secondary maximum of \hvezda\ has been
decreasing in brightness while the two minima have become nearly identical.
It should be noted that the time 
dependence
of parameters
$a_1,\ a_2$ has no obvious effect on the star's period.

Changes in the shape of light curves are very rare among CP stars: the only star
with well-documented light curve variations is the rapidly rotating silicon CP
star 56 Arietis.  The light curve variations are accompanied by a steady
lengthening of the stellar period \citep{zischw,admal}. On the other hand, in
the most rapidly braked CP star known (HD\,37776), we did not find any secular
changes in its light curve \citep{brzda}.  However, it is premature to speculate
further on the nature of long-term light-curve variations of \hvezda\ without
further photometric and spectroscopic monitoring.

To compare the observed and calculated light-curve shapes and phases, we
determined the shift between phases computed with our ephemeris and the
ephemeris used by \citet{leh2} for surface mapping.  The observed rotational
variations of the silicon line equivalent widths demonstrated that Lehmann's
phases precede ours by $0.1955(25)$. We have corrected for this phase
difference in our analysis of the elemental surface distributions.

\section{Calculation of the light curve}

\subsection{Stellar parameters} 

\begin{figure*}[t]
\centering
\resizebox{0.4\hsize}{!}{\includegraphics{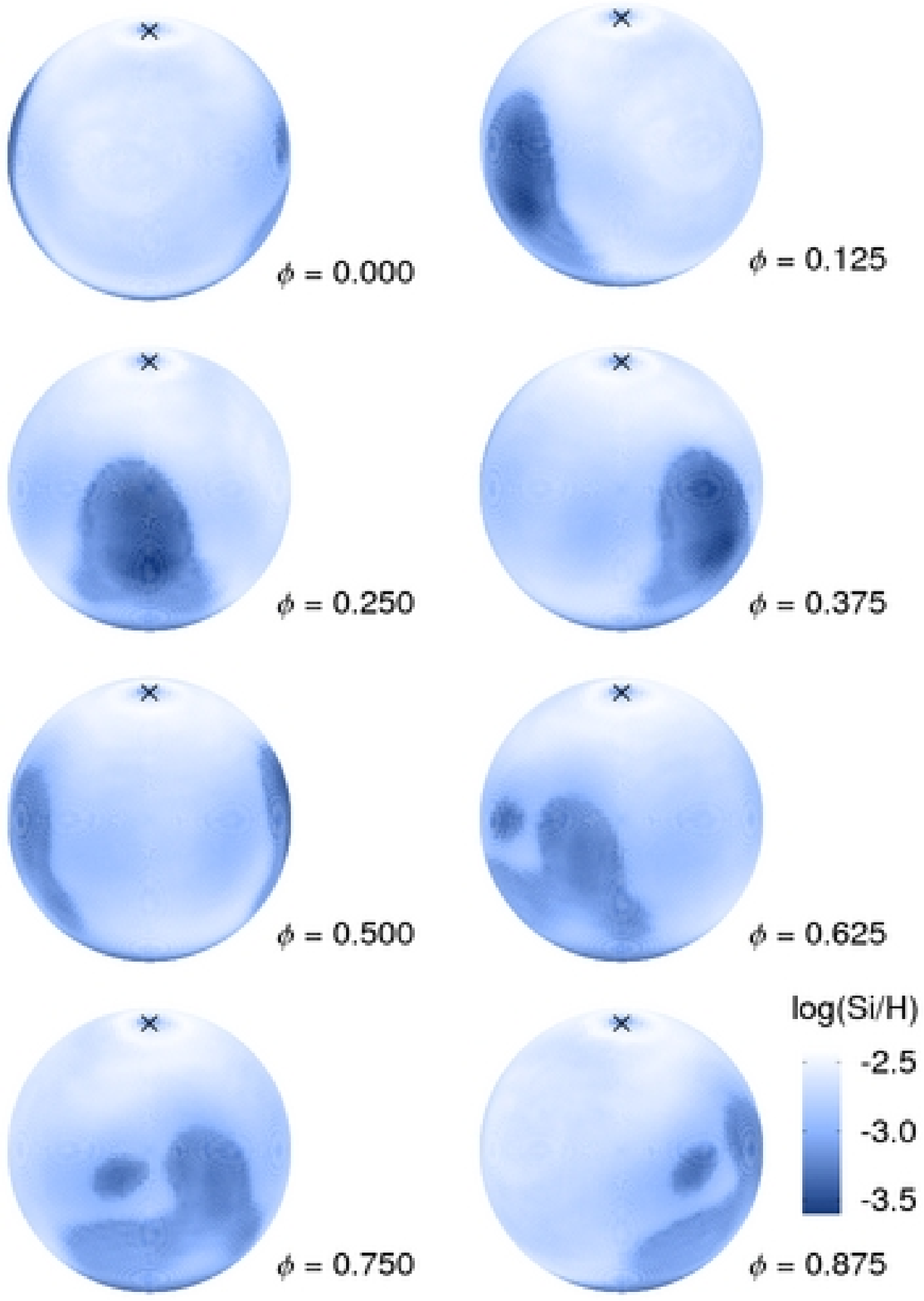}}
\resizebox{0.4\hsize}{!}{\includegraphics{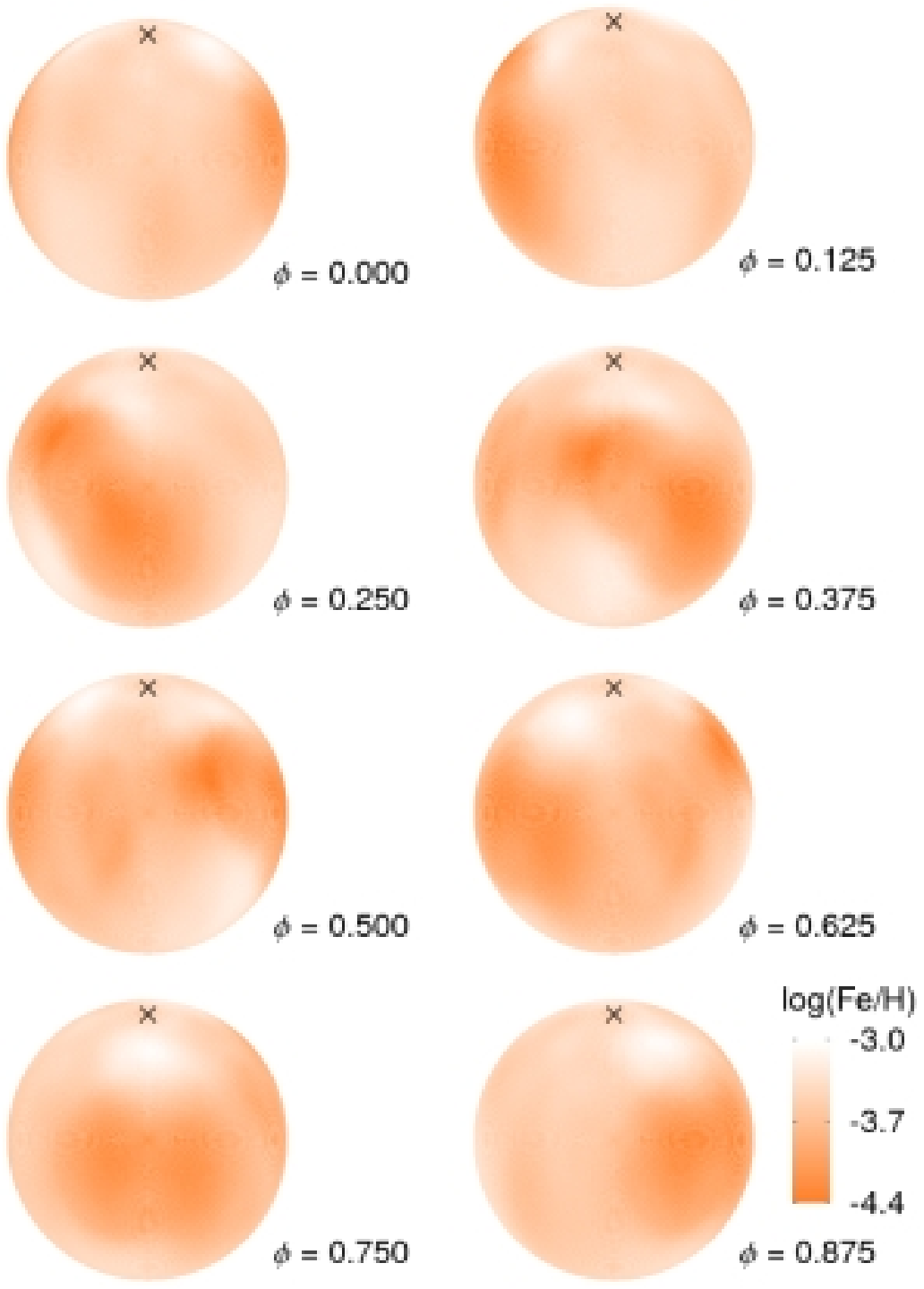}}
\caption{Observed distribution of silicon (left) and iron (right) on
the visible surface of \hvezda\ at different rotational phases \citep{leh2}.
Phases were calculated with the parameters given in
Eq.~\eqref{linef}.}
\label{sifepovrch}
\end{figure*}

The stellar parameters of \hvezda\ adopted in this study, as well as the
abundances of silicon and iron derived from the abundance maps
(see Fig.~\ref{sifepovrch}) and the abundance of helium are taken from
\citet[][see Table~\ref{hvezda}] {leh1,leh2}.
The abundances are relative to hydrogen,
$\varepsilon_\text{el}=\log\zav{N_\text{el}/N_\text{H}}$.
Note that the abundances used by \citet{leh1,leh2} are defined slightly
differently as $\log(N_\text{el}/N_\text{total})$.

\begin{table}[ht]
\caption{Stellar parameters of \hvezda.}
\label{hvezda}
\begin{center}
\begin{tabular}{lcc}
\hline
Effective temperature ${{T}_\mathrm{eff}}$ & ${14\,500}$\,K \\
Surface gravity ${\log g}$ (CGS) & ${4.2}$ \\
Inclination ${i}$ & ${65^\circ}$ \\
Helium abundance&$\varepsilon_\text{He}=-3.5$\\
Silicon abundance& $-3.6\lesssim\varepsilon_\text{Si}\lesssim-2.4$ \\
Iron abundance&$-4.4\lesssim\varepsilon_\text{Fe}\lesssim-3.0$  \\
\hline
\end{tabular}
\end{center}
\end{table}

\subsection{Model atmospheres and synthetic spectra}

We used the code TLUSTY \citep{tlusty,hublaj,hublad,lahub} for the model
atmosphere calculations. We computed plane-parallel model atmospheres in
LTE, taking into account the atomic data suitable for B type stars
\citep{bstar2006}. The atomic data for silicon are taken from \citet{mendo},
\citet{maslo93}, and Taylor (2008), in preparation; for iron from
\citet{kur22}, \citet{nah96}, \citet{nah97}, \citet{bau97}, and
\citet{bau96}, and for other elements from \citet{top1}, \citet{topf},
\citet{toptul}, \citet{topp}, \citet{toph}, and \citet{napra}.

We assumed fixed values of the effective temperature and surface gravity and
adopted models with a generic value $v_\text{turb}=2\,\text{km}\,\text{s}^{-1}$.
We assumed the number density of helium relative to hydrogen to be
$\log\zav{N_\text{He}/N_\text{H}}=-3.5$ \citep{leh2}. The abundance of silicon
and iron differed in individual models as explained below. We used the solar
abundance of other elements according to \citet{asgres}.

For the spectrum synthesis (from which we calculate the photometric colours), we
used the SYNSPEC code. We took into account the same transitions as for the
model atmosphere calculations, and we also included the lines of all elements
with the atomic number $Z\leq30$ not included in the model atmosphere
calculation. We computed angle-dependent intensities for $20$ equidistantly
spaced values of $\mu=\cos\theta$, where $\theta$ is the angle between the
normal to the surface and the line of sight.

The model atmospheres and angle-dependent intensities
$I(\lambda,\theta,\varepsilon_\text{Si},\varepsilon_\text{Fe})$ were
calculated for a two-parametric grid of silicon and iron abundances
$\varepsilon_\text{Si}=\log\zav{N_\text{Si}/N_\text{H}}= -4.0$, $-3.5$,
$-3.0$, $-2.5$, and $-2.0$, and
$\varepsilon_\text{Fe}=\log\zav{N_\text{Fe}/N_\text{H}}=-4.5$, $-4.0$,
$-3.5$, and $-3.0$.

\subsection{Phase dependent magnitude}
\label{vypocet}

The radiative flux observed at the distance $D$ from the star with
radius $R_*$ in a colour $c$ is calculated as

\begin{equation}
\label{vyptok}
f_c=\zav{\frac{R_*}{D}}^2\intvidpo I_c(\theta,\Omega)\cos\theta\,\text{d}\Omega,
\end{equation}
where the intensity $I_c(\theta,\Omega)$ at each surface point with
surface spherical coordinates $\Omega$ is obtained by means of
interpolation between the intensities\footnote{Note that the SYNSPEC
code outputs the specific intensities per unit frequency (not per unit
wavelength).} $I_c(\theta,\varepsilon_\text{Si},\varepsilon_\text{Fe})$
calculated from the grid of synthetic spectra as

\begin{equation}
I_c(\theta,\varepsilon_\text{Si},\varepsilon_\text{Fe})=
\int_0^{\infty}\Phi_c(\lambda) \,
I(\lambda,\theta,\varepsilon_\text{Si},\varepsilon_\text{Fe})\, \text{d}\lambda.
\end{equation}

The transmissivity function $\Phi_c(\lambda)$ of a given filter $c$
of the Str\"omgren and $\Delta a$ photometric systems is
approximated for simplicity by a Gauss function that peaks at the
central wavelength of the colour ${\lambda_c}$ with dispersion
${\sigma_c}$ (see Table~\ref{uvby}).
The values for the $uvby$ (Str\"omgren) photometric system are taken from
\citet{cox}. Values for the $\Delta a$ system (colours $g_1$ and $g_2$) are
taken from \citet{kupzen}.

\begin{table}[th]
\caption{Central wavelengths and dispersions of the Gauss filter
simulating the transmissivity functions.}
\label{uvby}
\begin{center}
\begin{tabular}{lcccccc}
\hline
Colour              & ${u}$& ${v}$& ${b}$& $g_1$ & $g_2$ & ${y}$\\
\hline
${\lambda_a}$ [\AA] & 3500 & 4100 & 4700 & 5000 & 5220 & 5500\\
${\sigma_a}$  [\AA] & 230  & 120  & 120  &  65  &  65  & 120 \\
\hline
\end{tabular}
\end{center}
\end{table}

The observed magnitude difference is
\begin{equation}
\label{velik}
\Delta m_{c}=-2.5\,\log\,\zav{\frac{{f_c}}{f_c^\mathrm{ref}}},
\end{equation}
where $f_c$ is calculated from Eq.~\eqref{vyptok} and
${f_c^\mathrm{ref}}$ is the reference flux obtained under the
condition that the mean magnitude over all phases is zero.

\section{Influence of abundance anomalies on emergent fluxes}
\label{kaptoky}

\begin{figure*}[tp]
{\centering \resizebox{0.49\hsize}{!}{\includegraphics{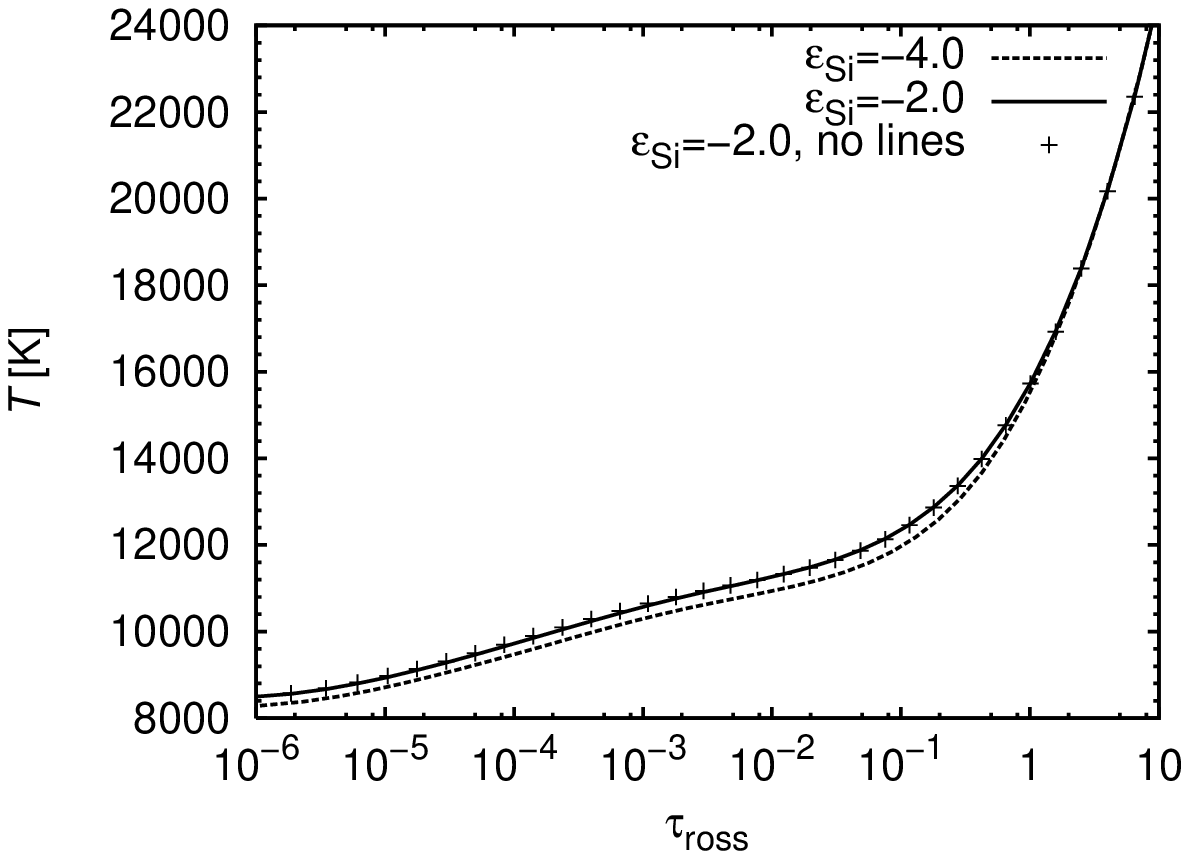}}}
\resizebox{0.49\hsize}{!}{\includegraphics{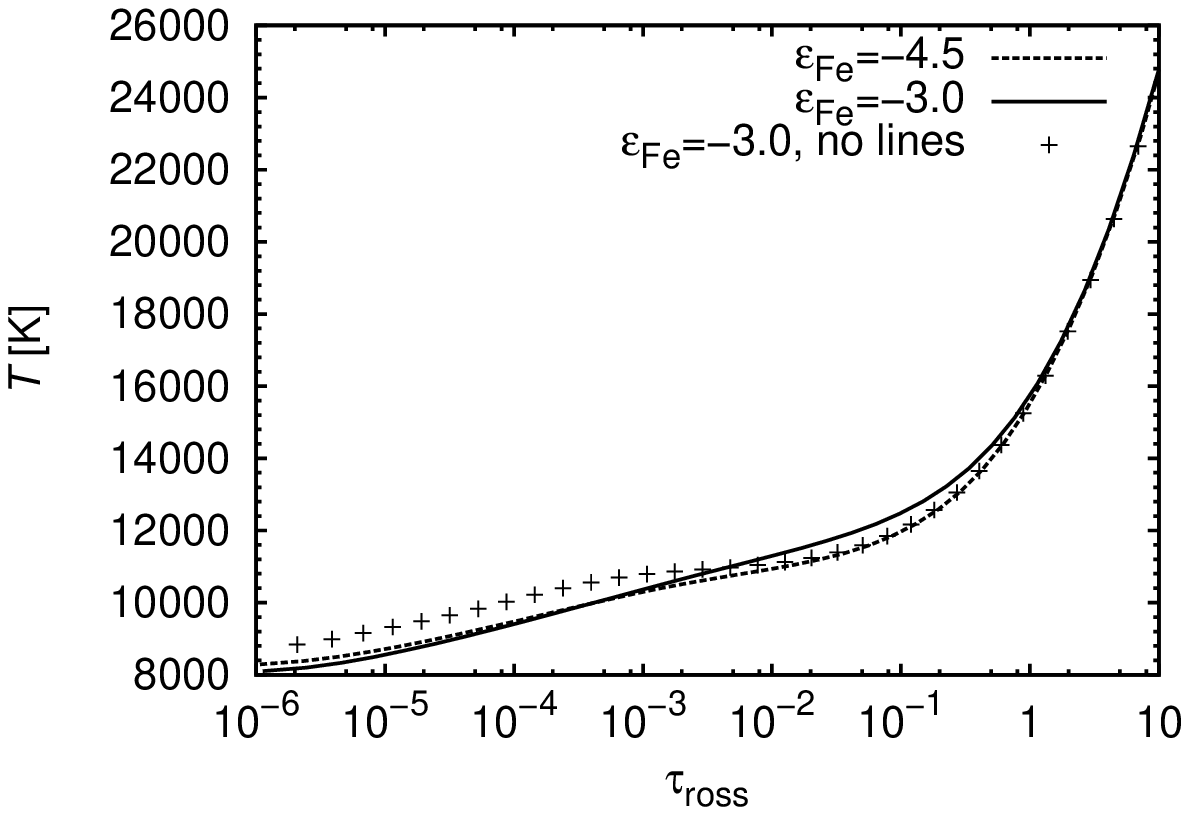}} \caption{The
dependence of temperature on the Rosseland optical depth
$\tau_\text{ross}$ in atmospheres with various chemical
compositions. {\em Left}: Influence of silicon abundance on the
temperature for $\varepsilon_\text{Fe}=-4.5$. {\em Right}: Influence
of iron abundance on the temperature for $\varepsilon_\text{Si}=-4.0$.
Crosses denote atmospheres with enhanced silicon or iron abundance, but
neglecting the opacity due to line transitions of these elements.}
\label{tep}
\end{figure*}

\begin{figure*}[tp]
\centering \resizebox{0.9\hsize}{!}{\includegraphics{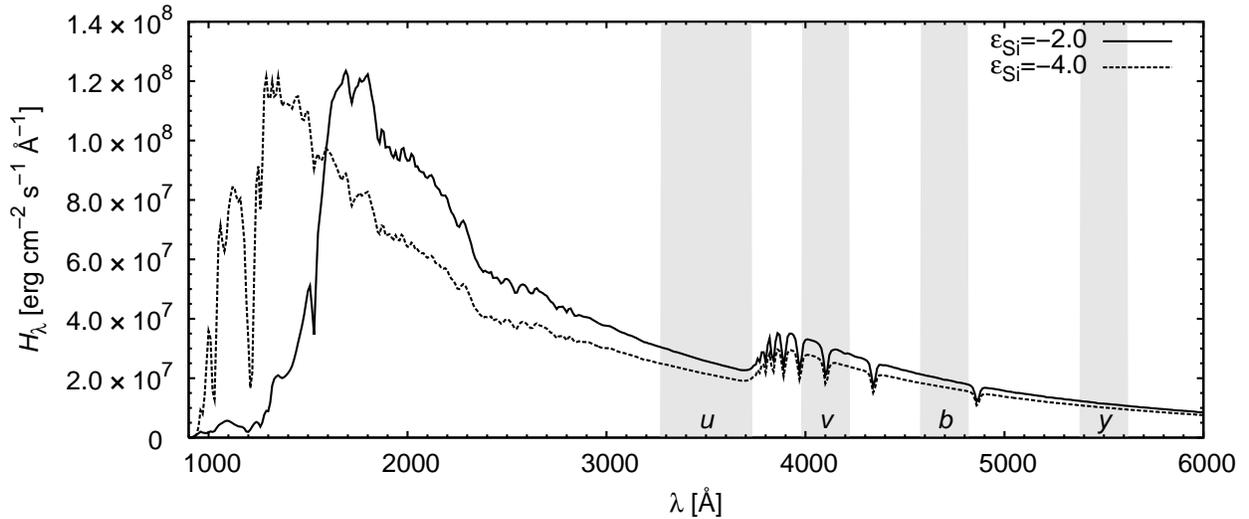}}
\caption{The emergent flux from atmospheres with different
silicon abundances. The flux was smoothed by a Gaussian filter with
a dispersion of $10\,\AA$ to show the changes in continuum, which
are important for photometric variability. The passbands of the
$uvby$ photometric system are also shown on the graph (gray areas).
The fluxes were calculated with TLUSTY for an iron abundance
$\varepsilon_\text{Fe}=-4.5$.}
\label{sitoky}
\end{figure*}

\begin{figure*}[tp]
\centering
\resizebox{0.9\hsize}{!}{\includegraphics{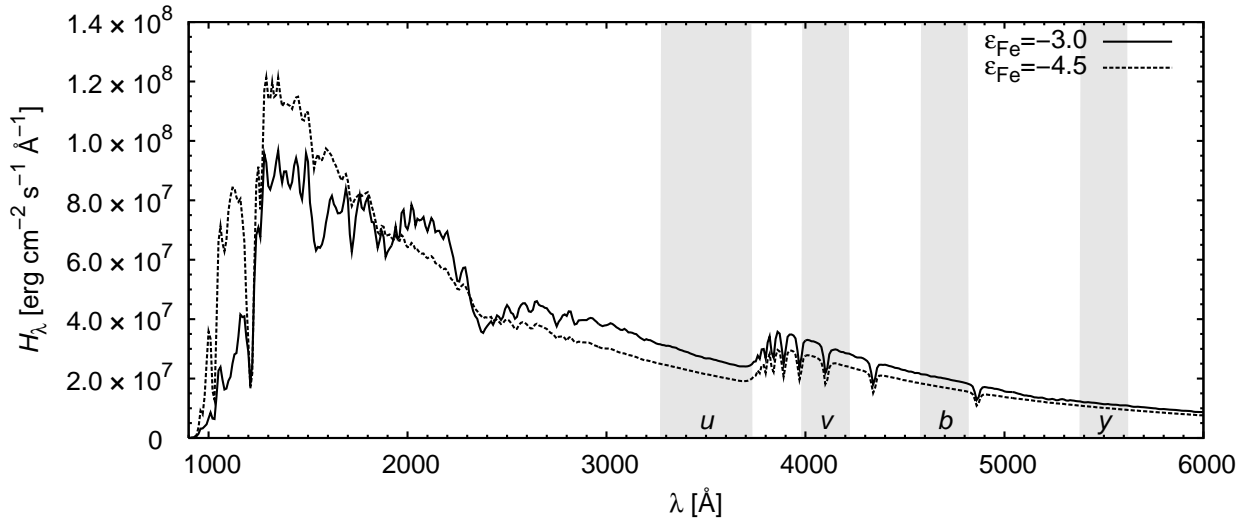}}\hfill
\caption{The same as Fig.~\ref{sitoky} but for iron. The fluxes were
calculated with TLUSTY for a silicon abundance
$\varepsilon_\text{Si}=-4.0$.}
\label{fetoky}
\end{figure*}

As shown in Fig.~\ref{tep}, the enhanced silicon abundance results in
the increase of the temperature in the continuum-forming region (within
$\tau_\text{ross}\approx0.1-1$) of the model atmosphere. The increased
temperature is caused by enhanced silicon opacity in the model atmosphere.
To understand the contribution of bound-free and bound-bound (line)
processes to the temperature increase, we calculated model atmospheres
with enhanced silicon abundance but neglecting silicon line transitions%
\footnote{Note that the resonance structure of silicon bound-free
transitions was included in this test.}
(see also Fig.~\ref{tep}). Neglecting silicon line transitions does not
significantly influence the atmospheric temperature; consequently,
enhanced opacity due to silicon bound-free transitions is the main cause
of the temperature increase.

The silicon bound-free transitions are important mainly in the
ultraviolet (UV) spectral regions at wavelengths shorter than
$1600\,$\AA. In atmospheres with overabundant silicon, the short-wavelength
part of the spectrum is redistributed to the longer wavelengths of
the UV spectrum, and also to the visible spectral regions
(see Fig.~\ref{sitoky}). Consequently, the silicon-rich spots are bright
in the $uvby$ colours, and are dark in the ultraviolet bands with
$\lambda\lesssim1600\,$\AA.

A similar situation occurs for iron overabundance. Enhanced iron
abundance leads to an increased temperature in the continuum-forming
region between $\tau_\text{ross}\approx1$ and
$\tau_\text{ross}\approx5\times10^{-3}$ (see Fig.~\ref{tep}). Unlike
silicon, enhanced iron abundance warms the atmosphere above
$\tau_\text{ross}\gtrsim 5 \times 10^{-3}$ mainly by line
transitions. On the other hand, the outermost regions
($\tau_\text{ross}\lesssim10^{-4}$) of the model with
$\varepsilon_\text{Fe}=-3$ are slightly
cooler than those with lower iron abundance. This effect is also due
to iron lines, as the temperature is higher in the model with
neglected line opacity (see also Fig.~\ref{tep}). A similar effect
was reported by \citet{preslo}.

The influence of iron overabundance on the spectrum is more complicated than the
influence of silicon overabundance, as shown in Fig.~\ref{fetoky}. There are
several depressions in the UV spectral region caused by numerous iron lines, but
the flux with $\lambda\gtrsim2500\,$\AA\ increases with increasing iron
abundance. Generally, the iron line transitions redistribute the emergent UV
radiation with $\lambda\lesssim1700\,$\AA\ primarily to the long wavelength part
of the spectrum with $\lambda\gtrsim2500\,$\AA. Consequently, iron-rich regions
are bright in the $uvby$ colours, whereas they are dark in the ultraviolet bands
with $\lambda\lesssim1700\,$\AA.

\begin{figure}[t]
\centering
\resizebox{0.9\hsize}{!}{\includegraphics{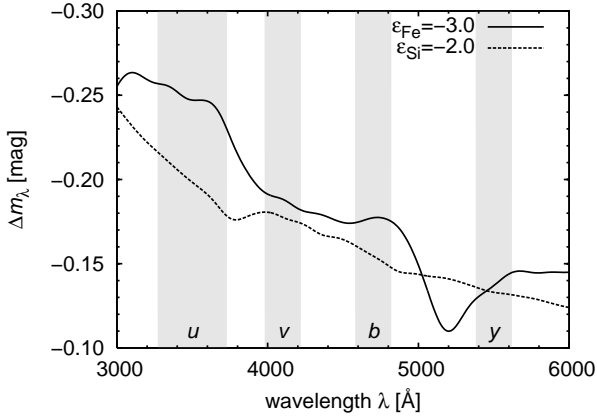}}
\caption{The magnitude difference $\Delta m_\lambda$ (see
Eq.~\eqref{tokmagroz}) between the emergent fluxes calculated with enhanced
abundance of either silicon or iron and fluxes calculated assuming
$\varepsilon_\text{Si}=-4$ and $\varepsilon_\text{Fe}=-4.5$.
The fluxes were heavily smoothed by a Gaussian filter with a dispersion
of $100\,$\AA. The depression at $5200\,$\AA\ due to numerous iron lines
is clearly apparent.}
\label{magtoky}
\end{figure}

These flux changes can be detected as a change in the apparent magnitude
of the star (see Fig.~\ref{magtoky}). Here we plot the relative magnitude
difference defined as
\begin{equation}
\label{tokmagroz}
\Delta m_\lambda=-2.5\log\zav{
  \frac{H_\lambda(\varepsilon_\text{Si},\varepsilon_\text{Fe})}
       {H_\lambda^\text{ref}}},
\end{equation}
where $H_\lambda^\text{ref}$ is the reference flux for
$\varepsilon_\text{Si}=-4.0$, $\varepsilon_\text{Fe}=-4.5$. For
both silicon and iron overabundance models, the absolute value
of the relative magnitude difference decreases with increasing
wavelength. For the silicon model, the decrease is nearly featureless,
whereas the iron model exhibits several depressions in $\Delta m_\lambda$
due to the cumulation of iron lines. The most prominent feature at about
$5200\,$\AA\ contributes significantly to the $5200\,$\AA\ depression
frequently observed in the spectra of CP stars. \citet{preslo} arrived
at the same conclusion.

\section{Predicted light variations}

\begin{figure}[t]
\centering
\resizebox{0.9\hsize}{!}{\includegraphics{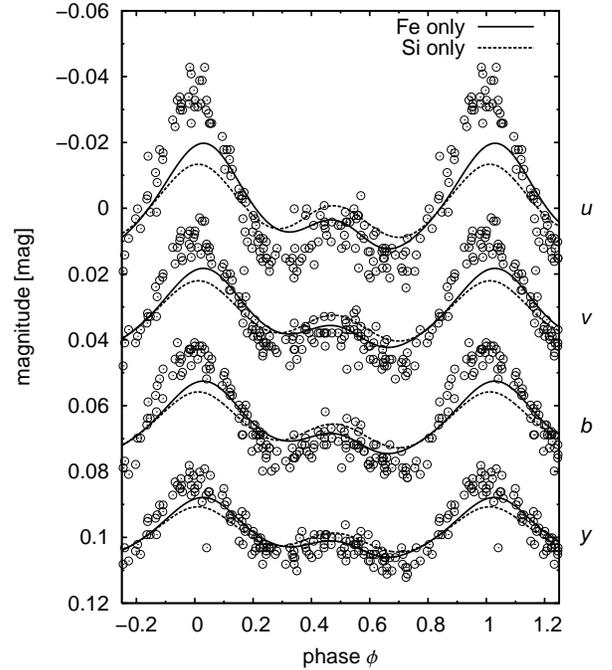}}
\caption{Light variations of \hvezda\ calculated from the silicon
abundance maps only (dashed line) and from the iron abundance maps
only (solid line).  For both cases, $\Lambda=0.0001$. Observed light
variations (open circles) are taken from \citet{adela}.}
\label{sife_hvvel}
\end{figure}

Predicted light curves are calculated from the surface abundance maps
derived by \citet{leh2} and from the emergent fluxes computed with the
SYNSPEC code, applying Eq.~\eqref{velik} for individual rotational phases.
Since \citet{leh2} provide surface maps with different regularisation
parameters $\Lambda$, we selected a map with $\Lambda=0.0001$, which is
the most detailed one, for our initial calculations.

To study the influence of silicon and iron separately, we first calculated the
light variations due to silicon only, assuming a constant iron abundance of
$\varepsilon_\text{Fe}=-4.5$. The observed light maximum occurs at the same
phase at which the silicon lines have their maximum strength ($\phi=0$). As
silicon rich regions are bright in visible bands, our predicted light maximum
should also occur at this phase. Indeed, there is a good agreement of both the
times of maxima and minima of our predicted light curve and the observed light
curve from \citet{adela}, though the amplitudes are different
(Fig.~\ref{sife_hvvel}).

A similar test was performed using
only
the iron abundance map. 
The silicon
abundance was assumed to be $\varepsilon_\text{Si}=-4.0$. The iron lines are
also observed to have their maximum strength close to phase $\phi=0$, i.e.,
during observed light maximum. As the iron-rich regions are bright in the
$uvby$ colours (see Fig.~\ref{fetoky}), the predicted light curve due to
iron abundance variations alone also has a light maximum at phase $\phi=0$,
in agreement with the observations. The predicted light curves due to iron
only, displayed in Fig.~\ref{sife_hvvel}, also have lower amplitudes than
the observed ones.

\begin{figure}[t]
\centering
\resizebox{0.9\hsize}{!}{\includegraphics{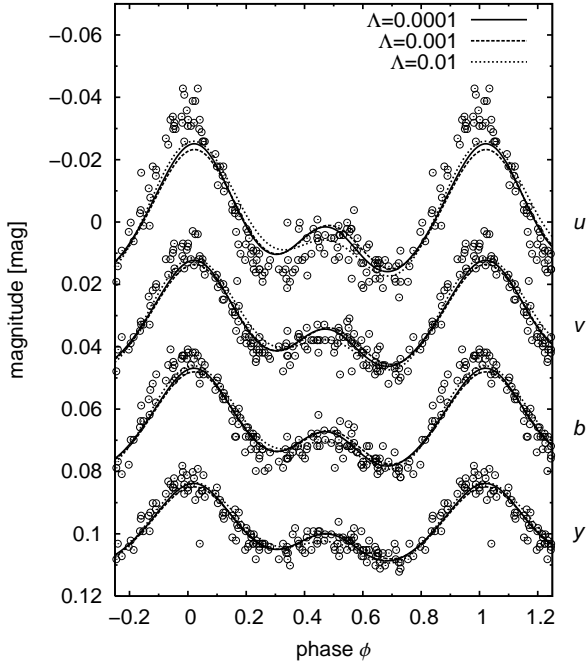}}
\caption{Predicted light variations of \hvezda\ computed
taking into account both the silicon and iron surface abundance distributions
derived by \citet{leh2} for different regularisation parameters $\Lambda$.
Observed light variations (circles) are taken from \citet{adela}.}
\label{hr7224_hvvel}
\end{figure}

\begin{figure}[t]
\centering
\resizebox{0.9\hsize}{!}{\includegraphics{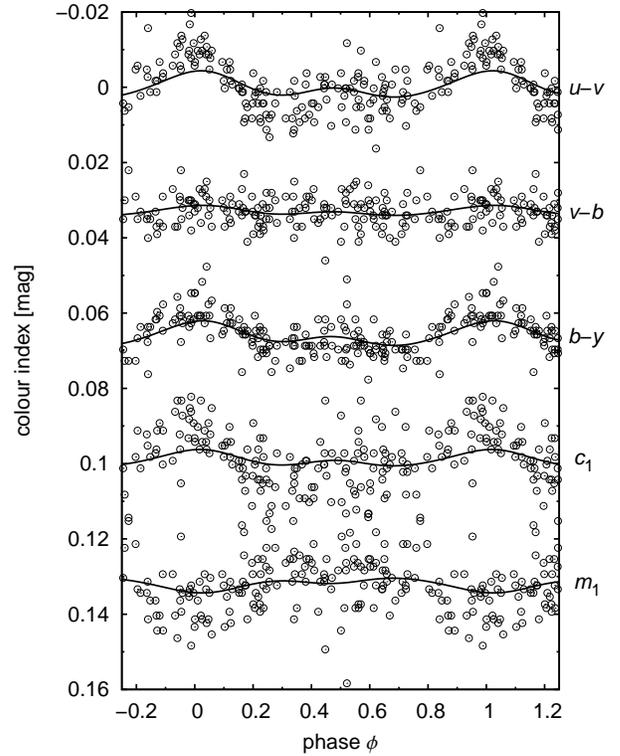}}
\caption{Predicted variations of colour indexes (solid line) calculated
from the surface abundance maps derived with $\Lambda=0.0001$ compared
with the observations.}
\label{hr7224_uvby}
\end{figure}

Including the surface distribution of both silicon and iron in the calculation
of the light curves (Fig.~\ref{hr7224_hvvel}), we obtain very good agreement
between the observed and predicted light curves in the $v$, $b$ and $y$
wavelength bands. In the $u$ band, however, the observed maximum is higher and
the first minimum is deeper than predicted. A similar situation is seen in the
colour indexes: $(v-b)$ and $(b-y)$ agree well with the observations while the
$(u-b)$ curve is not a good fit, due to the differences in the $u$ band fit
(Fig.~\ref{hr7224_uvby}). The observed Balmer discontinuity index $c_{\rm 1}$
exhibits large scatter due to the combination of the noisy $u$, $v$ and $b$
light curves, but its variation is nonetheless in accordance with the
observations of \citet{leh2} on a possible variability of H$\beta$ line.
Similarly, the $m_{\rm 1}$ index, which reveals the effect of the metallic
bound-bound absorption in the violet when compared with the $b$ and $y$ regions,
shows a phase-dependence. The differences between the predicted curves of
$c_{\rm 1}$ and $m_{\rm 1}$ and the observed ones clearly point to the existence
of an additional, unknown mechanism working in the violet band that still has to
be investigated (Fig.~\ref{hr7224_hvvel}, and see also Sect.~\ref{kapsrov}).

\begin{figure}[t]
\centering
\resizebox{0.8\hsize}{!}{\includegraphics{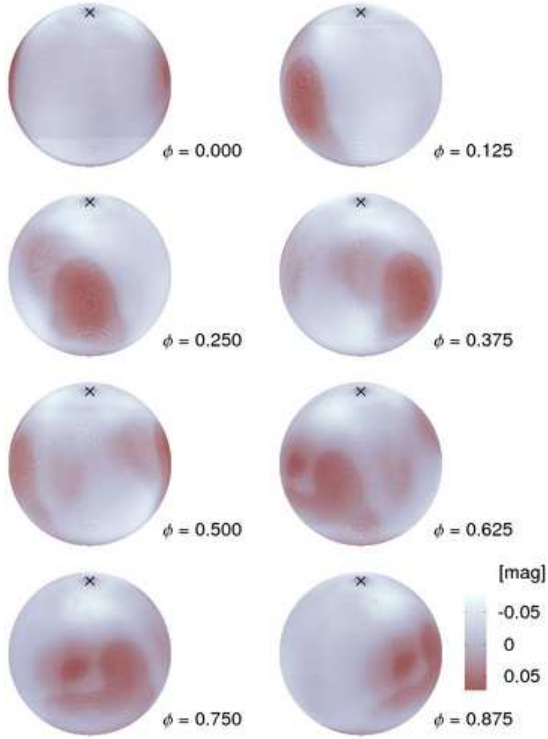}}
\caption{The emergent intensity (in the $y$ band, $\mu=1$)
from individual surface elements of \hvezda\ at various rotational phases.}
\label{hr7224_povrch}
\end{figure}

Thus, the inhomogeneous surface distribution of silicon and iron revealed in
spectroscopic observations results in the appearance of spots on the stellar
surface causing the photometric variability (see Fig.~\ref{hr7224_povrch}).

\section{Discussion}

\subsection{Detailed comparison of observed and predicted light curves}
\label{kapsrov}


\begin{figure}[t]
\centering \resizebox{0.9\hsize}{!}{\includegraphics{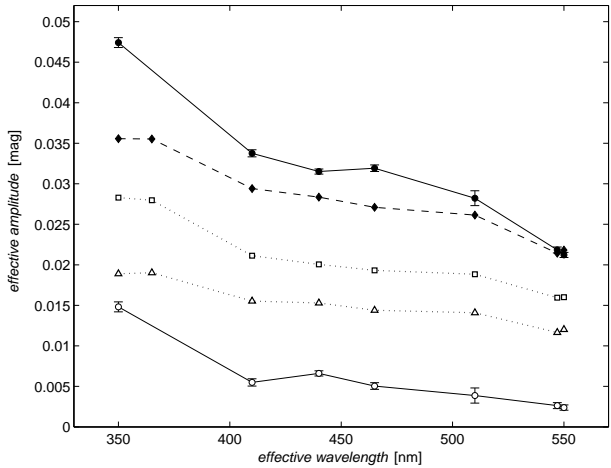}}
\caption{The effective amplitudes of the predicted and observed light
curves plotted versus the effective wavelengths of the various bands.
The observed light-curve amplitudes are plotted with filled circles
($\bullet$). The predicted amplitudes calculated using both silicon and
iron maps are filled diamonds ($\blacklozenge$), amplitudes calculated
using iron only are open squares, ($\Box$), silicon only are
triangles ($\triangle$), and finally the residuals between
observation and theory as open circles ($\circ$).
Observed amplitudes in $B_{\text{T}}V_{\text{T}}$ and $U$ are not plotted
here due to their large uncertainty.}
\label{amplitudy}
\end{figure}

The predicted and observed light curves differ slightly in their shapes
(Fig.~\ref{hr7224_hvvel}) and amplitudes (Fig.~\ref{amplitudy}).  The small
differences in the observed vs. calculated amplitudes, leads to the differences
in the observed vs. calculated colour indexes, especially for $c_1$ and $m_1$
(see Fig.~\ref{hr7224_uvby}). The difference between the observed and predicted
light curves in Fig.~\ref{hr7224_hvvel} could be explained by the presence of an
additional photometric spot(s) on \hvezda\, caused by overabundance of other
element(s). We suspect chromium or magnesium. \citet{preslo} suggested chromium
as an element capable of significant influence over the emergent flux from a
star. On the other hand, \citet{leh1} found variability in the magnesium lines
of \hvezda\ with maximum strength occurring roughly at the same phase as the
maximum difference between theory and observation in Fig.~\ref{hr7224_hvvel}
(slightly before the maximum strength of silicon lines).  This, along with the
fact that the calculated effect of the magnesium abundance on the emergent flux
has a significantly larger amplitude in the $u$ band rather than in $v$, $b$ and
$y$, argues for magnesium as the cause of the residuals between observed and
calculated light curves when considering only silicon and iron. Note however,
that the maximum magnesium abundance derived by \citet{leh1} without surface
mapping is too small to affect the spectral energy distribution significantly.
The same is also true for helium and oxygen.

\citet{leh2} derived three abundance maps calculated with a different
regularisation parameter, $\Lambda$, corresponding to a different assumed
minimum size of the surface inhomogeneities.  The shapes of the light curves
calculated using the abundance maps obtained with a different value of $\Lambda$
are similar (see Fig.~\ref{hr7224_hvvel}). A detailed comparison of these curves
shows that better agreement between theory and observation is obtained for the
most complex surface map, i.e., for $\Lambda=0.0001$. However, the difference
between observed and predicted light curves is significantly larger than the
difference among the individual predicted curves calculated with different
values of $\Lambda$.

\subsection{Model assumptions}

There are several effects that can influence the predicted light curves
\citep[see][~for a more detailed discussion]{myhd37776}. For example, NLTE
effects may influence the continuum flux distribution, but since \hvezda\
surface maps were derived using LTE models, we confine ourselves to LTE models
only. Strong surface magnetic field also influences the emergent spectral energy
distribution \citep[e.g.,][]{malablat}. However, as all available measurements
of surface magnetic field are negative and the corresponding upper limit is too
low to signicantly influence the emergent flux, we neglect the influence of
magnetic field.

\hvezda\ is classified as a helium-weak chemically peculiar star, so a question
could arise concerning the effect of the uneven surface distribution of helium
on the light curve. In \citet{myhd37776} we showed that helium influences the
spectral energy distribution only in the case when it is overabundant (with
respect to solar), so we neglect any possible inhomogeneous helium surface
distribution in our model. We also tested how accumulation of He in the
sub-photospheric layers affects emergent flux and draw a conclusion that for
$\tau_\text{ross} > 1$ this has no effect on the light variability.

\subsection{UV variations}

The redistribution of the flux from short to longer wavelengths
is one of the consequences of the proposed mechanism for light
variability in \hvezda.  From our calculated model fluxes in
Figs.~\ref{sitoky}, \ref{fetoky}, we predict that the star should have
a flux minimum in the short-wavelength $\lambda\lesssim1600\,$\AA\ part
of the spectrum during the visible light maximum and vice versa. A
similar antiphase behaviour of the short-wavelength UV and optical
light curves has been reported for other CP stars \citep[e.g.,][]{sokold}.

Unfortunately, there is no UV light curve of \hvezda\ available.
However, the star was observed in UV by the TD1 spacecraft
\citep{dzemar1,dzemar2}. We note that our fluxes calculated with
overabundant iron in Fig.~\ref{fetoky} are in qualitative agreement
with those observed by TD1 at different phases. In accordance with
observations, the maximum UV light amplitude occurs at about
$1400\,$\AA, and the variations at this wavelength are
anticorrelated with the variations in the visible. The variations at
$1600-1900\,$\AA\ are very small, and the variations at
$2100-2300\,$\AA\ are inversely correlated with that at $1400\,$\AA.
The absence of observed light variations at $2740\,$\AA\ may be
connected with strong iron lines in this spectral region. Finally,
the observed amplitude of the light variations at $1400\,$\AA\
\citep[which is $0.24\,$mag,][]{dzemar2} is in very good agreement
with those derived from our models ($0.21\,$mag for a Gaussian
filter with dispersion $50\,$\AA).

\subsection{Variations of $a$ }

The peculiarity index $a$ introduced by \citet{acko} as
\begin{equation}
a=g_2-(g_1+y)/2
\end{equation}
is based on the existence of flux depressions \citep{kodaidep}
correlated with the chemical peculiarity of a given star. To test
whether the peculiarity demonstrated in the $a$-index is indeed
caused by numerous iron lines \citep{preslo}, we calculated the
peculiarity index $a$ for individual model atmospheres used in our
previous calculations.

\begin{figure}[t]
\centering
\resizebox{0.9\hsize}{!}{\includegraphics{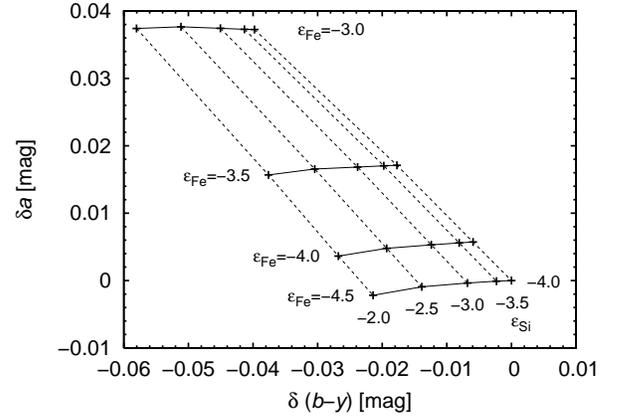}}
\caption{The dependence of the $\delta a$ index on $\delta(b-y)$ for
individual models from the model grid. The $\delta a$ values for
models with the same iron abundance are connected with solid lines, and
$\delta a$ values for the models with the same silicon composition are
connected with dashed lines.}
\label{abyhvvelsit}
\end{figure}

\begin{figure}[t]
\centering
\resizebox{0.9\hsize}{!}{\includegraphics{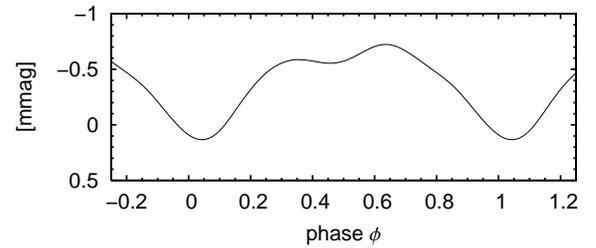}}
\caption{The light curve in $\delta a$ index}
\label{abyhvvel}
\end{figure}

The dependence of $a$ on the iron and silicon abundance can be seen
in Fig.~\ref{abyhvvelsit}, where we plot the dependence of $\delta
a$ on $\delta(b-y)$. These indices are defined relative to their
values for the model with $\varepsilon_\text{Si}=-4.0$ and
$\varepsilon_\text{Fe}=-4.5$,
\begin{align}
\delta a =& a-a_{\varepsilon_\text{Si}=-4.0,\varepsilon_\text{Fe}=-4.5},\\
\delta(b-y)=&(b-y)-(b-y)_{\varepsilon_\text{Si}=-4.0,\varepsilon_\text{Fe}=-4.5}.
\end{align}
With increasing iron abundance, the $\delta a$ index increases, whereas
the influence of silicon on $\delta a$ is only marginal. On the other hand,
both the silicon and iron abundances influence the value of the $\delta(b-y)$
index. This implies that for a star with a given effective temperature (and
an assumed homogeneous surface distribution of elements), it is possible
to infer the value of iron and silicon abundance directly from photometry.
One has to keep in mind, however, that other elements may also influence
the value of $\delta(b-y)$ and $\delta a$ \citep[cf.,][]{kupzenil,
preslo}. Consequently, such procedure is likely possible only for hot CP stars.
Finally,
for stars with a strong magnetic field, the Zeeman splitting of iron lines
may also influence the $a$ index \citep{malablat}.

As the star rotates, surface regions with different chemical composition appear
on the visible disc and consequently $\delta a$ and $\delta(b-y)$ also vary with
rotational phase (see Figs.~\ref{hr7224_uvby}, \ref{abyhvvel}). Here we plot the
light curve calculated as described in Sect.~\ref{vypocet} with filter
parameters taken from Tab.~\ref{uvby}. From this plot, we can conclude that the
amplitudes of $\delta a$ variations are very small, on the order of
millimagnitudes.

\subsection{H$\beta$ line profile variations}

\begin{figure}[t]
\centering
\resizebox{0.9\hsize}{!}{\includegraphics{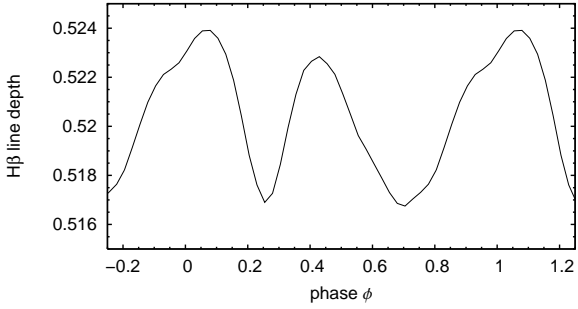}}
\caption{The calculated variations of H$\beta$ line depth with phase}
\label{hbeta}
\end{figure}

The fact that we are able to predict the correct amplitudes of brightness
variations in both the UV and visual spectral domains from just the abundance
maps of iron and silicon supports the conclusion that the effective temperature
is constant on the surface of \hvezda. This finding weakens one of possible
explanations of Balmer lines' profile variations as summarised by \citet{leh2}.
Since the effect of the Lorenz force on the line profiles is mainly on the line
wings \citep[c.f.,][]{denvid}, it cannot explain the line profile variability
observed in \hvezda. We tested if our model is also able to explain the observed
H$\beta$ line variability. For this purpose we calculated the line profiles in
individual phases using the code developed by \citet{skaldip}. The code
numerically integrates the emergent intensity $I_c(\theta,\Omega)$ over the
visible stellar surface $\Omega$ taking into account the Doppler shift due to
the stellar rotation. The emergent intensity from each surface point is
calculated by interpolating the synthetic spectra
$I(\lambda,\theta,\varepsilon_\text{Si},\varepsilon_\text{Fe})$ for appropriate
surface abundance distribution of silicon and iron \citep{leh2}. We assume fixed
effective temperature and surface gravity. The predicted phases of the maximum
line depth (Fig.~\ref{hbeta}) roughly correspond to the observed ones
\citep[Fig.~(13) therein]{leh1}, also the shape of the minimum at the phase
$\phi\approx0.2$ is roughly correct, however the observed amplitude of the
minimum at the phase $\phi\approx0.6$ seems to be by a factor of about 2 higher.
We conclude that either another elements has to play the role in the H$\beta$
line variability, or the observed difference between observation and theory is
partially caused by somehow noisy observational data.

\section{Conclusions}

We successfully simulated the light variability of \hvezda\ directly
from the silicon and iron surface abundance maps derived by
\citet{leh2}. There is very good agreement between the observed
and predicted light variability in the $uvby$ bands of the Str\"omgren
photometric system. We did not introduce any free parameter to improve
the agreement between the theoretical and observed light curves.

The rotationally modulated light variability of \hvezda\ is caused by
the flux redistribution due to iron line transitions, silicon bound-free
transitions, and by the inhomogeneous surface distribution of these
elements. This picture is also supported by the agreement between the
predicted behaviour of the UV flux distribution and that observed
by the TD1 satellite.

We support the conclusion of \citet{preslo} that numerous iron lines contribute
significantly to the well-known depression at $5200\,$\AA. Moreover, we show
that iron is able to influence the peculiarity index $a$, whereas silicon's
influence is marginal. With a suitable calibration, this could enable the
derivation of abundances in hot CP stars directly from photometry.

We conclude that a promising explanation for the light variations in CP stars is
a flux redistribution through line and bound-free transitions combined with the
inhomogeneous surface distribution of various elements.

\begin{acknowledgements}
We thank Dr. A.~Tkachenko for providing us with surface maps and
Drs.~E.~Paunzen and D.~Shulyak for the discussion of this topic.
This work was supported by grants GA\,\v{C}R 205/06/0217, GA\,\v{C}R
205/08/H005, VEGA 2/6036/6, MEB 080832/SK-CZ-0090-07, MEB 060807,
and partly by GA\,\v{C}R 205/07/0031.
This research made use of NASA's Astrophysics Data System, the
SIMBAD database, operated at the CDS, Strasbourg, France and the on-line
database of photometric observations of mCP stars \citep{mikdata}.
GWH acknowledges support from NASA, NSF, Tennessee State University, and
the State of Tennessee through its Centers of Excellence program.
\end{acknowledgements}


\begin{thebibliography}{}
\bibitem[Adelman(1997)]{adela} Adelman, S.\,J. 1997, \aap, 122, 249
\bibitem[Adelman(2004)]{adelc} Adelman, S.\,J. 2004, MNRAS, 351, 823
\bibitem[Adelman et al.(2001)]{admal} Adelman, S. J., Malanushenko,
    V., Ryabchikova, T. A., \& Savanov, I. 2001, \aap, 375, 982
\bibitem[Asplund et al.(2005)]{asgres}
        Asplund, M., Grevesse, N., \& Sauval, A. J. 2005,
        Cosmic Abundances as Records of Stellar Evolution and
        Nucleosynthesis, ASP Conf. Ser. 336, eds. T. G. Barnes III,
        F. N. Bash (San Francisco: ASP), 25
\bibitem[Bautista(1996)]{bau96} Bautista, M. A. 1996, A\&AS, 119, 105
\bibitem[Bautista \& Pradhan(1997)]{bau97} Bautista, M. A., \& Pradhan, A. K.
    1997, A\&AS, 126, 365
\bibitem[Bohlender et al.(1993)]{bolek} Bohlender, D. A., Landstreet, J. D.,
    \& Thompson, I. B. 1993, A\&A, 413, 273
\bibitem[Butler et al.(1993)]{maslo93} Butler, K., Mendoza, C., \& Zeippen,
    C. J. 1993, J. Phys. B, 26, 4409
\bibitem[Cowley et al.(1969)]{cow} Cowley, A., Cowley, C., Jaschek,
    M., et al. 1969 AJ, 74, 375
\bibitem[Cox(2000)]{cox} Cox, A.\,N., ed., 2000, Astrophysical Quantities (New
    York: AIP Press)
\bibitem[Eaton, Henry, \& Fekel(2003)]{ehf03}
    Eaton, J. A., Henry, G. W., \& Fekel, F. C. 2003, in The Future of
    Small Telescopes in the New Millennium, Volume II - The Telescopes
    We Use, ed. T. D. Oswalt (Dordrecht:  Kluwer), 189
\bibitem[ESA(1997)]{esa97} ESA 1997, The Hipparcos and Tycho Catalogs, SP--1200
\bibitem[ESA(1998)]{esa98} ESA 1998, The Hipparcos and Tycho Catalogs, Celestia
    2000, SP--1220
\bibitem[Fernley et al.(1999)]{topf} Fernley, J. A., Hibbert, A., Kingston,
    A. E., \& Seaton, M. J. 1999, J. Phys. B, 32, 5507
\bibitem[Henry(1995a)]{h95a}
        Henry, G.W. 1995a, in ASP Conf. Ser. 79, Robotic Telescopes: Current
        Capabilities, Present Developments, and Future Prospects for Automated
        Astronomy, ed. G. W. Henry \& J. A. Eaton (San Francisco: ASP), 37
\bibitem[Henry(1995b)]{h95b}
        Henry, G. W. 1995b, in ASP Conf. Ser. 79, Robotic Telescopes: Current
        Capabilities, Present Developments, and Future Prospects for Automated
        Astronomy, ed. G. W. Henry \& J. A. Eaton (San Francisco: ASP), 44
\bibitem[Hibbert \& Scott(1994)]{toph} Hibbert, A., \& Scott, M. P. 1994,
    J. Phys. B, 27, 1315
\bibitem[Hubeny(1988)]{tlusty} Hubeny, I. 1988, Comput. Phys. Commun., 52, 103
\bibitem[Hubeny \& Lanz(1992)]{hublaj} Hubeny, I., \& Lanz, T. 1992, A\&A, 262, 501
\bibitem[Hubeny \& Lanz(1995)]{hublad} Hubeny, I., \& Lanz, T. 1995, ApJ, 439, 875
\bibitem[Jamar(1977)]{dzemar1} Jamar, C. 1977, A\&A, 56, 413
\bibitem[Jamar(1978)]{dzemar2} Jamar, C. 1978, A\&A, 70, 379
\bibitem[Khan \& Shulyak(2006)]{malablat} Khan, S.\,A., \& Shulyak, D.\,V.
     2006, A\&A, 454, 933
\bibitem[Khan \& Shulyak(2007)]{preslo} Khan, S.\,A., \& Shulyak, D.\,V. 2007,
     A\&A, 469, 1083
\bibitem[Khokhlova et al.(2000)]{choch} Khokhlova, V.\,L., Vasilchenko, D.\,V.,
        Stepanov, V.\,V., \& Romanyuk, I.\,I. 2000, AstL, 26,  177
\bibitem[Kochukhov et al.(2005)]{malablaj} Kochukhov, O., Khan, S., \&
        Shulyak, D. 2005, A\&A, 433, 671
\bibitem[Kodaira(1967)]{koda} Kodaira, K. 1967, Ann. Tokyo Astr. Obs., 10, 157
\bibitem[Kodaira(1969)]{kodaidep} Kodaira, K. 1969, ApJL, 157, 59
\bibitem[Krivosheina et al.(1980)]{krivo} Krivosheina, A.\,A., Ryabchikova,
     T.\,A., \& Khokhlova, V.\,L. 1980, Nauchnye Informatsii, Ser. Astrof.,
     43, 70
\bibitem[Krti\v{c}ka et al.(2007)]{myhd37776} Krti\v{c}ka, J., Mikul\'a\v sek,
    Z., Zverko, J., \& Zi\v z\v novsk\'y, J. 2007, A\&A, 470, 1089
\bibitem[Kupka et al.(2003)]{kupzen} Kupka, F., Paunzen, E., \& Maitzen, H. M.
    2003, MNRAS, 341, 849
\bibitem[Kupka et al.(2004)]{kupzenil} Kupka, F., Paunzen, E., Iliev, I. Kh.,
     \& Maitzen, H. M. 2003, MNRAS, 352, 863
\bibitem[Kurucz(1994)]{kur22} Kurucz, R. L. 1994, Kurucz CD-ROM 22, Atomic Data
    for Fe and Ni (Cambridge: SAO)
\bibitem[Landstreet \& Borra(1978)]{labor} Landstreet, J. D. \& Borra, E. F.
    1978, ApJL, 224, 5
\bibitem[Lanz et al.(1996)]{lanko} Lanz, T., Artru, M.-C., Le Dourneuf, M., \&
    Hubeny, I. 1996, A\&A, 309, 218
\bibitem[Lanz \& Hubeny(2003)]{lahub} Lanz, T., \& Hubeny, I. 2003, ApJS, 146,
    417
\bibitem[Lanz \& Hubeny(2007)]{bstar2006} Lanz, T., \& Hubeny, I. 2007, ApJS,
    169, 83
\bibitem[Lehmann et al.(2006)]{leh1} Lehmann, H., Tsymbal, V.,
    Mkrtichian, D. E., \& Fraga, L. 2006, A\&A, 457, 1033
\bibitem[Lehmann et al.(2007)]{leh2} Lehmann, H., Tkachenko, A., Fraga, L.,
    Tsymbal, V., \& Mkrtichian, D. E. 2007, A\&A, 471, 941
\bibitem[Luo \& Pradhan(1989)]{top1} Luo, D., \& Pradhan, A. K., 1989,
    J. Phys. B, 22, 3377
\bibitem[Maitzen(1976)]{acko} Maitzen, H. M. 1976, A\&A, 51, 223
\bibitem[Mendoza et al.(1995)]{mendo} Mendoza, C., Eissner, W., Le Dourneuf,
    M., \& Zeippen, C. J. 1995, J. Phys. B, 28, 3485
\bibitem[Michaud(2005)]{mpoprad} Michaud, G. 2005, in The A-Star
        Puzzle, IAU Symposium No. 224, eds. J. Zverko,
        J. \v{Z}i\v{z}\v{n}ovsk\'y, S. J. Adelman, \& W.\,W. Weiss
        (Cambridge: Cambridge Univ. Press),  173
\bibitem[Mikul\'a\v sek et al.(2007a)]{mikdata} Mikul\'a\v sek, Z.,
    Jan\'\i k, J., Zverko, J., et al. 2007a, Astron.~Nachr., 328, 10
\bibitem[Mikul\' a\v sek et al.(2007b)]{mikzoo} Mikul\'a\v sek, Z.,
    Krti\v cka, J., Zverko, J., et al. 2007b, in Physics of
    Magnetic Stars, ed. I. I. Romanyuk \& D. O. Kudryavtsev
(SAO, Nizhnij Arkhyz), 300
\bibitem[Mikul\'a\v sek et al.(2008a)]{simply} Mikul\'a\v sek,
    Z., Gr\'af, T., Krti\v{c}ka, J., et al. 2008a, CAOSP 38, 363
\bibitem[Mikul\'a\v sek et al.(2008b)]{brzda} Mikul\'a\v sek,
    Z., Krti\v{c}ka, J., Henry, G. W. et al. 2008b, A\&A, 485, 585
\bibitem[Molnar(1972)]{mlyn} Molnar, M.\,R. 1972, ApJ, 175, 453
\bibitem[Molnar(1973)]{molnar} Molnar, M.\,R. 1973, ApJ, 179, 527
\bibitem[Nahar(1996)]{nah96} Nahar, S. N. 1996, Phys. Rev. A, 53, 1545
\bibitem[Nahar(1997)]{nah97} Nahar, S. N. 1997, Phys. Rev. A, 55, 1980
\bibitem[Nahar \& Pradhan(1993)]{napra} Nahar, S. N., \& Pradhan, A. K.
    1993, J. Phys. B 26, 1109
\bibitem[Nakajima(1985)]{nakaji} Nakajima, R. 1985, Ap\&SS, 116, 285
\bibitem[Peach et al.(1988)]{topp} Peach, G., Saraph, H. E., \& Seaton, M. J.
    1988, J. Phys. B, 21, 3669
\bibitem[Peterson(1970)]{peter} Peterson, D.\,M. 1970, ApJ, 161, 685
\bibitem[Seaton et al.(1992)]{topt} Seaton, M.\,J., Zeippen, C.\,J., Tully,
     J.\,A., et al. 1992, Rev. Mexicana Astron. Astrofis., 23, 19
\bibitem[Shaham(1986)]{kuba} Shaham, J. 1986, ApJ, 310, 780
\bibitem[Shulyak et al.(2007)]{denvid} Shulyak, D., Valyavin, G., Kochukhov,
O. et al. 2007, A\&A, 464, 108
\bibitem[Skalick\'y(2008)]{skaldip} Skalick\'y, J. 2008, diploma thesis, Masaryk
University, Brno
\bibitem[Smith \& Groote(2001)]{smigro} Smith, M.\,A., \& Groote, D. 2001,
     A\&A, 372, 208
\bibitem[Sokolov(2006)]{sokold} Sokolov, N.\,A. 2006, MNRAS, 373, 666
\bibitem[Townsend et al.(2005)]{towog} Townsend, R.\,H.\,D. Owocki, S.\,P., \&
        Groote D. 2005, ApJ, 630, L81
\bibitem[Tully et al.(1990)]{toptul} Tully, J. A., Seaton, M. J., \&
    Berrington, K. A. 1990, J. Phys. B, 23, 3811
\bibitem[Vauclair(2003)]{vaupreh} Vauclair, S. 2003, Ap\&SS, 284, 205
\bibitem[Weiss et al.(1976)]{biltep} Weiss, W. W., Albrecht, R., \& Wieder, R.
    1976, A\&A, 47, 423
\bibitem[Winzer(1974)]{winzer} Winzer, J. E. 1974, Ph.D. Thesis, Univ. Toronto
\bibitem[\v{Z}i\v{z}\v{n}ovsk\'{y} et al.(2000)]{zischw}
    \v{Z}i\v{z}\v{n}ovsk\'{y}, J., Schwartz, P., \& Zverko, J. 2000,
    IBVS, 4835
\end{thebibliography}
\end{document}